\documentclass[preprintnumbers,amsmath,amssymb]{revtex4}
\usepackage{graphicx}
\usepackage{amsmath}
\usepackage[utf8]{inputenc}
\usepackage[T1]{fontenc}

\begin{document}
\draft
\title{Equilibrium frame reveals hidden {PT} symmetry of passive systems}

\author{Grzegorz Chimczak}
\affiliation{Institute of Spintronics and Quantum Information, Faculty of Physics, Adam Mickiewicz University, 61-614 Pozna\'{n}, Poland}
\email{chimczak@amu.edu.pl}
\author{Anna Kowalewska-Kud{\l}aszyk}
\affiliation{Institute of Spintronics and Quantum Information, Faculty of Physics, Adam Mickiewicz University, 61-614 Pozna\'{n}, Poland}
\author{Ewelina Lange}
\affiliation{Institute of Spintronics and Quantum Information, Faculty of Physics, Adam Mickiewicz University, 61-614 Pozna\'{n}, Poland}
\author{Karol Bartkiewicz}
\affiliation{Institute of Spintronics and Quantum Information, Faculty of Physics, Adam Mickiewicz University, 61-614 Pozna\'{n}, Poland}
\affiliation{RCPTM, Joint Laboratory of Optics of Palacký University and Institute of Physics of Czech Academy of Sciences, 17. listopadu 12, 771 46 Olomouc, Czech Republic}

\begin{abstract}
We discuss how introducing an equilibrium frame, in which a given Hamiltonian has balanced loss and gain terms, can reveal ${\cal{PT}}$ symmetry hidden in non-Hermitian Hamiltonians of dissipative systems. Passive ${\cal{PT}}$-symmetric Hamiltonians, in which only loss is present and gain is absent, can also display exceptional points, just like ${\cal{PT}}$-symmetric systems, and therefore are extensively investigated. We demonstrate that non-Hermitian Hamiltonians, which can be divided into a ${\cal{PT}}$-symmetric term and a term commuting with the Hamiltonian, possess hidden ${\cal{PT}}$ symmetries. These symmetries become apparent in the equilibrium frame. We also show that the number of eigenstates having the same value in an exceptional point is usually smaller in the initial frame than in the equilibrium frame. This property is associated with the second part of the Hamiltonian.
\end{abstract}

\maketitle

\section{Introduction}
In recent years, there has been an increasing interest in exploring non-Hermitian physical systems as a source of novel physical effects~\cite{ozdemir2019parity,el2018non,Miri2019}. 
It has been shown that the special group of non-Hermitian Hamiltonians, which possess parity-time (${\cal{PT}}$) symmetry, can exhibit entirely real spectra, like Hermitian Hamiltonians~\cite{Bender_98,bender2007making,BenderBook}. 

From both theoretical and experimental points of view, special attention is paid to the degeneracies induced by PT symmetries. Such degeneracies, known as exceptional points (EPs), are the points in a parametric space, at which  eigenvalues and eigenvectors coincide. Only non-Hermitian systems can display EPs~\cite{Mostafazadeh15,Minganti_20}. At EPs a ${\cal{PT}}$ phase transition occurs and all interesting physics associated with the enhancement of nonclassical system's features can be observed.

There are numerous examples  of such novel features detected directly at and around EPs in atomic \cite{Zhang_16, Wang_16}, optical \cite{Miri2019}, optomechanical \cite{Jing_17, Djorwe_19}, plasmonic \cite{Alaeian_14, Kuo_20, Park_20}, photonic crystalline \cite{Cerjan_16,Chen_19}, and many other physical systems. Among others, one can find examples of enhancement of weak signal sensing \cite{Chen_17}, enhancement of spontaneous emission \cite{Lin_16}, asymmetric light propagation \cite{Markis_08,Jin17}, single-mode laser \cite{Feng_14}, electromagnetically induced transparency \cite{Wang_20} just to name the few. Therefore, there is an increasing interest among researchers to look for new physical systems in which EPs existence can be confirmed \cite{wrona2020}.

Special attention in the research on EPs and ${\cal{PT}}$ symmetries is put to optical and photonic systems \cite{ozdemir2019parity, Miri2019, Wang16PRA}. Their usefulness comes from the natural presence of both gain and loss of energy processes in the evolution. Therefore, finding the right balance between gain and loss is crucial for the non-Hermitian system to be able to observe the effects connected to EPs.

The EPs are usually studied in semiclassical regime, where optical and photonic systems make use of strong classical external fields. However, there are also some attempts in which the fully quantum description of the system's evolution is applied \cite{Minganti_19, Perina_19, Arkhipov_20b}.
Moreover, it has been shown that also systems without gain components, i.e., including only losses, can have EPs and exhibit the enhancement of linear and nonlinear interactions \cite{ozdemir2019parity, Arkhipov_20a}. When only losses are included, a system is known to have passive ${\cal{PT}}$ symmetry. Eigenenergies for such systems have a common imaginary part and its presence is not an obstacle in observing EPs. The first experimental realisations of passive PT systems were performed with the use of coupled waveguides \cite{Guo_09}. Then, other experiments with loss-induced PT symmetry breaking, for example, in optical resonators \cite{Peng_14}, externally modulating metamaterials \cite{Feng_13}  were performed. To reveal the desirable symmetry, the appropriate relation between the lossy components of both coupled modes has to be obtained. The result of that interplay is the increase of transmitted power in one of the  modes despite the fact that only lossy mechanisms are included. Surprisingly, the emergence of a slowly decaying mode is not contingent on the existence of EPs in systems where only lossy mechanisms are included~\cite{Joglekar18}.

In our considerations, we deal with the passive type of non-Hermitian systems. 
Our main aim is to show that if the whole non-Hermitian Hamiltonian, being not PT-symmetric, can be expressed as a composition of two parts: (i) standard PT-symmetric term and (ii) a term commuting with (i), then the hidden PT symmetry is present in this non-Hermitian Hamiltonian. We show that transforming the whole initial Hamiltonian to time-dependent state vector scaling reveals the hidden PT symmetry of the considered system and the presence of EPs is possible even though the whole system is not directly PT-symmetric. We will refer to the frame, where the hidden PT symmetry is clearly seen, as to equilibrium frame (EF).

We believe that our findings can be helpful in the investigation of novel physical systems in which the hidden presence of PT symmetry can be revealed by expressing a Hamiltonian, defined in an initial frame (IF), in EF.

\section{Results}

\subsection{Equilibrium Frame}
First, let us present the main idea of a transformation to EF. It is based on another one, frequently used in quantum optics,  transformation to a rotating frame. We assume that the total Hamiltonian can be written as a sum of two terms $H^{\cal{PT}}$ and $H_{0}$. The Schr\"odinger equation is thus given by ($\hbar=1$)
\begin{eqnarray}
  \label{eq:EF01}
  i\partial_t|\psi\rangle &=& ( H^{\cal{PT}} + H_{0} ) |\psi\rangle\, .
\end{eqnarray}
Now we make the substitution $|\psi\rangle=S\, |\widetilde{\psi}\rangle$, where $S$ and $|\widetilde{\psi}\rangle$ are time-dependent. If we set $S=\exp(-i\, H_{0}\, t)$ then the Schr\"odinger equation reduces to 
\begin{eqnarray}
  \label{eq:EF02}
  i\partial_t |\widetilde{\psi}\rangle &=& \widetilde{H} |\widetilde{\psi}\rangle\, ,
\end{eqnarray}
where $\widetilde{H} = S^{-1} H^{\cal{PT}} S$. In the case of the transformation to a rotating frame, $S$ is unitary, because $H_{0}$ is Hermitian. However, in the case of the transformation to EF the operator $S$ is not a unitary one, because $H_{0}$ is not Hermitian. In both cases $\widetilde{H}$ and $H^{\cal{PT}}$ have the same eigenvalues. In order to obtain a ${\cal{PT}}$-symmetric Hamiltonian in EF, we restrict ourselves to the cases, where $[H^{\cal{PT}},H_{0}]=0$. Using the Baker–Hausdorf lemma
\begin{eqnarray}
  \label{eq:EF03}
  e^Y X e^{-Y} &=& X+[Y,X]+(1/2!)\big[Y,[Y,X]\big]+\dots
\end{eqnarray}
one can easily prove that $\widetilde{H}=H^{\cal{PT}}$ for these cases.

For $[H^{\cal{PT}},H_{0}]=0$, both Hamiltonians have the same set of eigenstates, and then we may relate the eigenvalues of the Hamiltonian given in IF to those in EF. Therefore, an $i$-th eigenvalue of the total Hamiltonian in IF
\begin{eqnarray}
  \label{eq:EF04}
  H |\phi_{i}\rangle &=&  H^{\cal{PT}} |\phi_{i}\rangle + H_{0} |\phi_{i}\rangle 
  = \widetilde{H} |\phi_{i}\rangle + H_{0} |\phi_{i}\rangle \nonumber\\
  E_{i} |\phi_{i}\rangle &=& \widetilde{E}_{i} |\phi_{i}\rangle + E^{(0)}_{i} |\phi_{i}\rangle
\end{eqnarray}
 is equal to the sum of the $i$-th eigenvalue of $\widetilde{H}$ and the corresponding eigenvalue of $H_{0}$. This fact is important when one is looking for Hamiltonians displaying EPs, i.e., points in the parameter space, where two (or more) eigenvalues have the same value. If $H^{\cal{PT}}$ is PT-symmetric, then the Hamiltonian in EF, i.e., $\widetilde{H}$, can display EPs. If it is the case, then two eigenvalues of $\widetilde{H}$ have the same value (  $\widetilde{E}_{i}=\widetilde{E}_{j}$ for some $i$ and $j$). 
 Therefore, one may state that the Hamiltonian given in IF, i.e., $H=H^{\cal{PT}}+H_{0}$, being
 not a PT-symmetric one, can also display EP if $E^{(0)}_{i}=E^{(0)}_{j}$. 
 Then the second condition for EP will be also fulfilled, because $H^{\cal{PT}}$ and $H$ 
 have the same set of eigenstates. Therefore, if in this point the eigenvectors of $H^{\cal{PT}}$ coincide then in this point eigenvectors of $H$ also coincide. Thus, we can say that EF reveals the hidden symmetry of $H$. 

Eigenvalues of $H_{0}$ determine whether
$H$ displays EPs or not. Moreover, they determine the type of frame. If the eigenvalues of $H_{0}$ are real, then we have a transformation to a rotating frame. If they are imaginary, then we have a transformation to a frame, in which the eigenstates scale with time. In the case when a system is in unbroken, ${\cal{PT}}$-symmetric phase, i.e., $H^{\cal{PT}}$ has a real spectrum, and eigenvalues of $H_{0}$ are imaginary then we can assign physical meanings to these two parts of $H$: $H^{\cal{PT}}$ is the energy observable of the system and $H_{0}$ is a geometric part, which depends on the geometric nature of the Hilbert space~\cite{Zhang19a,Mostafazadeh18,Zhang19b}.

It is worth to note that the condition $[H^{\cal{PT}},H_{0}]=0$ does not mean that $H_{0}$ is a constant of motion, since $H^{\cal{PT}}$ is not Hermitian. The conserved quantities in ${\cal{PT}}$-symmetric Hamiltonian evolutions are given by intertwining operators~\cite{Bian20_c,MOSTAFAZADEH10_pH}.

It is also worth to mention that the equilibrium frame can also be useful to reveal hidden pseudo-Hermiticity of non-Hermitian Hamiltonians. It is known that ${\cal{PT}}$ symmetry is a special case of pseudo-Hermiticity~\cite{Mostafazadeh02_1,Mostafazadeh02_5,Mostafazadeh02_8}. If the total Hamiltonian can be written as a sum of two parts: pseudo-Hermitian part and $H_{0}$ commuting with the first part, then one can expect that the eigenvalues of the Hamiltonian given in IF are related to those in EF.

In the next sections, we are going to concentrate on three quantum systems described by non-Hermitian Hamiltonians to investigate the effect of the geometric part of $H$ on displaying EPs.

\subsection{Exceptional point in the simplest passive system}
\begin{figure}[ht]
    \centering
    \includegraphics[width=0.95\linewidth]{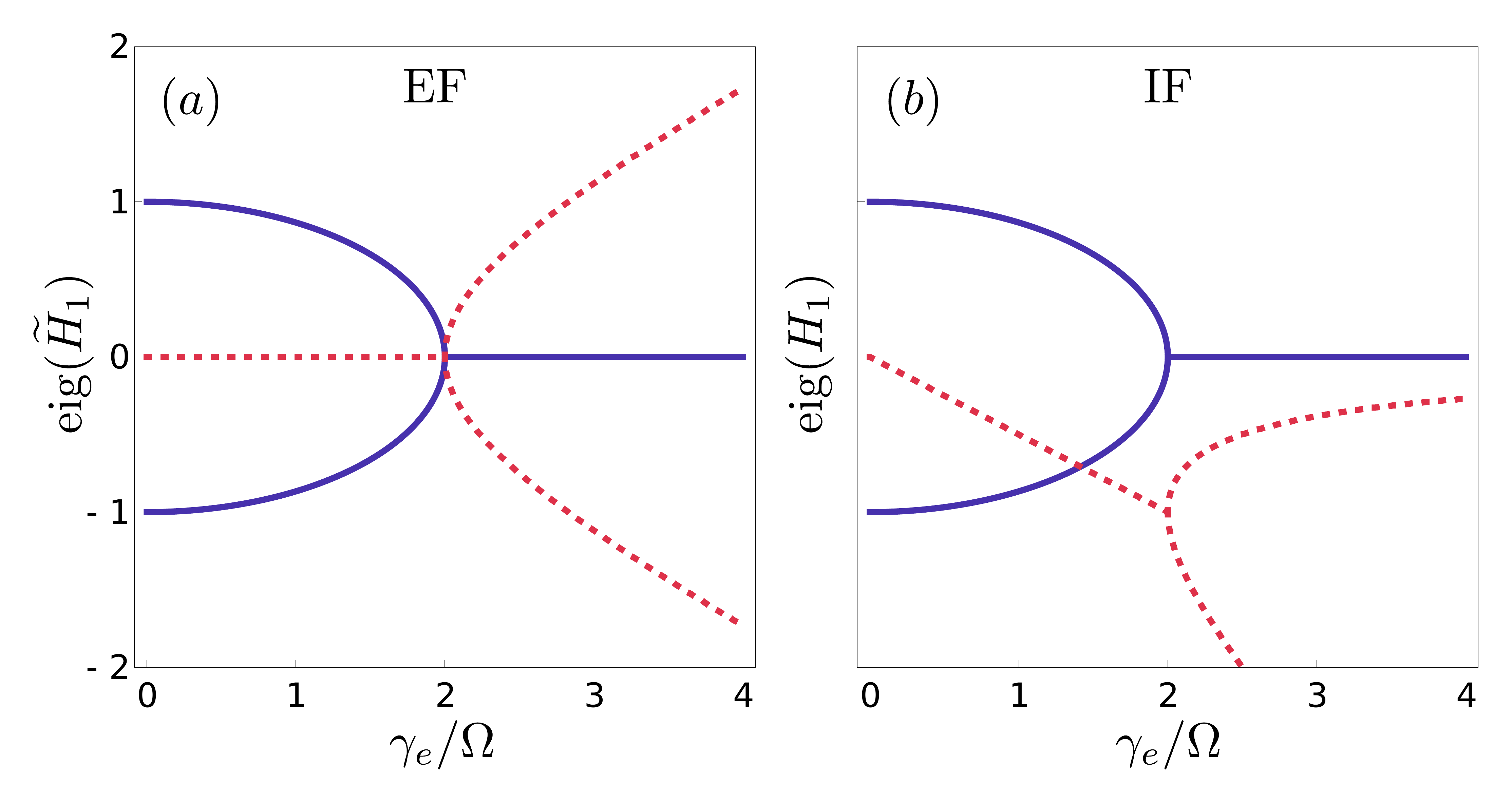}
    \caption{Real (solid lines) and imaginary (dashed lines) parts of two eigenvalues of the Hamiltonian~(\ref{eq:HamiltonianPT001}) expressed in (a) an equilibrium frame (EF), and (b) in an initial frame (IF). In EF the Hamiltonian is ${\cal{PT}}$-symmetric and it has an exceptional point in $\gamma_{e}/\Omega=2$, where both eigenvalues are the same. Below the exceptional point the eigenvalues are real, which is what is expected for a ${\cal{PT}}$-symmetric theory. In the initial frame this Hamiltonian also displays the exceptional point at $\gamma_{e}/\Omega=2$ despite the fact that it is not ${\cal{PT}}$-symmetric in this frame. Thus, equilibrium frames can reveal PT symmetry hidden in open systems with losses, but without gain.}
    \label{fig:passive-pt}
\end{figure}

The simplest system, where  hidden symmetry and an exceptional point can be found is a two-level atom driven by a classical laser field. 
In a rotating frame, the Hamiltonian that describes interaction of the atom with the laser field is given by
\begin{eqnarray}
  \label{eq:HamiltonianPT001}
  H_1&=& \Omega (\sigma_{eg} + \sigma_{ge}) - i\gamma_{e} \sigma_{ee}\, ,
\end{eqnarray}
where $\sigma_{ij}=|i\rangle\langle j|$, $|g\rangle$ is the atomic ground state, $|e\rangle$ is the excited state, $\Omega$ is an atom - classical laser field coupling strength, and $\gamma_{e}$ is an atomic polarization decay rate. Since $\sigma_{gg}+\sigma_{ee}=\hat{I}$ we can rewrite Eq.~(\ref{eq:HamiltonianPT001}) as
\begin{eqnarray}
  \label{eq:HamiltonianPT002}
  H_1&=& \Omega (\sigma_{eg} + \sigma_{ge}) - i\frac{\gamma_{e}}{2} \sigma_{ee} 
  +i\frac{\gamma_{e}}{2} \sigma_{gg} -i\frac{\gamma_{e}}{2} \hat{I}
\end{eqnarray}
A similar non-Hermitian Hamiltonian for a two-level spin model but with time-varying coupling constances was also considered in \cite{bagchi2018evolution} in which it was shown that for such a system it is possible to find a closed form of the evolution operator. 
Hamiltonian (\ref{eq:HamiltonianPT002}) can be divided into two parts $H_1=H^{\cal{PT}}_{1}+H^{(0)}_{1}$, where
\begin{eqnarray}
  \label{eq:HamiltonianPT003}
  H^{\cal{PT}}_{1}&=& \Omega (\sigma_{eg} + \sigma_{ge}) - i\frac{\gamma_{e}}{2} \sigma_{ee} 
  +i\frac{\gamma_{e}}{2} \sigma_{gg}\, , \nonumber\\
  H^{(0)}_{1}&=&-i\frac{\gamma_{e}}{2} \hat{I}\, .
\end{eqnarray}
We define the parity operator by the Pauli operator ${\cal{P}}=\sigma_{x}=\sigma_{eg} + \sigma_{ge}$~\cite{ozdemir2019parity,bender2007making,BenderBook} and ${\cal{T}}$ is a complex-conjugation operator (${\cal{T}} i {\cal{T}}=-i$). Using these definitions it is easy to check that $({\cal{PT}})H^{\cal{PT}}_{1}({\cal{PT}})=H^{\cal{PT}}_{1}$, i.e., $H^{\cal{PT}}_{1}$ is 
${\cal{PT}}$-symmetric. Note that $H^{(0)}_{1}$ is just an identity operator multiplied by an imaginary valued constant, and therefore, eigenvalues of $H_1$ differ from the corresponding eigenvalues of $H^{\cal{PT}}_{1}$ only by this imaginary constant. One can see that the geometric part of $H_1$, i.e., $H^{(0)}_{1}$, has no effect on EP. If $H^{\cal{PT}}_{1}$ reveals EP at some point of the parameter space, which means that the two eigenvalues of $H^{\cal{PT}}_{1}$ has the same real and imaginary parts 
and the corresponding eigenvectors coincide, then $H_1$ also reveals EP at this point of the parameter space as is seen in Fig.~\ref{fig:passive-pt}.
One can also see that the equilibrium frame, in which the Hamiltonian is given by $\widetilde{H}_{1}=H^{\cal{PT}}_{1}$, can reveal ${\cal{PT}}$ symmetry hidden in a non-Hermitian Hamiltonian. Therefore, transformation to the equilibrium frame can be useful in finding systems displaying EPs.

\subsection{Infinite-dimensional passive system}
Now let us investigate a more interesting case --- two electromagnetic modes coupled to each other, with damping present in both modes. The Hamiltonian which governs the evolution of this system is given by
\begin{eqnarray}
  \label{eq:HamiltonianPT01}
  H_2&=& g (a^{\dagger}b + b^{\dagger}a)-i\gamma_{a} a^{\dagger}a - i\gamma_{b} b^{\dagger}b\, ,
\end{eqnarray}
where $g$ is a coupling strength, $a$ and $b$ denote the annihilation operators, $\gamma_{a}$ and $\gamma_{b}$ are the field damping rates of both modes. This system is infinite-dimensional and it has infinitely many eigenvalues. We will compare the chosen eigenvalues of $H_2$ with corresponding eigenvalues of $\widetilde{H_2}$, i.e., we will compare the presented Hamiltonian, given in a initial and in a equilibrium frames.

Introducing $\kappa = (\gamma_{a}-\gamma_{b})/2$ and $\gamma = (\gamma_{a}+\gamma_{b})/2$ we can rewrite Eq.~(\ref{eq:HamiltonianPT01}) as 
\begin{eqnarray}
  \label{eq:HamiltonianPT011}
  H_2&=&g (a^{\dagger}b + b^{\dagger}a)-i\kappa a^{\dagger}a + i\kappa b^{\dagger}b 
  - i\gamma(a^{\dagger}a + b^{\dagger}b)
\end{eqnarray}
The Hamiltonian~(\ref{eq:HamiltonianPT011}) is not ${\cal{PT}}$-symmetric but, as we will see, equilibrium frame reveals the ${\cal{PT}}$ symmetry hidden in it. 
Having Hamiltonian (\ref{eq:HamiltonianPT011}), we consider two-mode open systems, where the two modes interact with each other. One mode, represented by the annihilation operator $a$, experiences damping while the other, represented by the annihilation operator $b$, experiences gain. In order to: prove that the Hamiltonian has a real secular equation, and be able to interpret ${\cal{P}}$ as a space reflection, we define it here in the following way
\begin{eqnarray}
  \label{eq:def_P}
{\cal{P}}&=&P_{\rm{S}} \exp[i\pi(a^{\dagger}a+b^{\dagger}b)]\, ,
\end{eqnarray}
where $P_{\rm{S}}$ is the exchange operator~\cite{HorodeckiPRL02}, which interchanges the modes spatially (i.e., $a\leftrightarrow b$). A matrix representation of $P_{\rm{S}}$ is given by a perfect shuffle~\cite{Loan00}.
We define the time-reversal operator ${\cal{T}}$ just as complex-conjugation operator (${\cal{T}} i {\cal{T}}=-i$). 

Note that ${\cal{P}}$ given by Eq.~(\ref{eq:def_P}) is a reflection operator (i.e., ${\cal{P}}={\cal{P}}^{-1}$) and $[{\cal{P}},{\cal{T}}]=0$. Using it and formulas: $\exp(\alpha a^{\dagger} a) a \exp(-\alpha a^{\dagger} a) = \exp(-\alpha) a$
and $\exp(\alpha a^{\dagger} a) a^{\dagger} \exp(-\alpha a^{\dagger} a) = \exp(\alpha) a^{\dagger}$, one can easily check that 
$({\cal{PT}}) a ({\cal{PT}}) = -b$, $({\cal{PT}}) a^{\dagger}({\cal{PT}}) = -b^{\dagger}$, $({\cal{PT}}) b ({\cal{PT}}) = -a$, $({\cal{PT}}) b^{\dagger}({\cal{PT}}) = -a^{\dagger}$ and $({\cal{PT}}) i ({\cal{PT}}) = -i$.

One can check that $H_2$ is not ${\cal{PT}}$-symmetric by applying the above-mentioned symmetry transformation to the bosonic field.

Hamiltonian~(\ref{eq:HamiltonianPT011}) can also be written as a sum of two parts 
\begin{eqnarray}
  \label{eq:HamiltonianPT02}
  H^{\cal{PT}}_{2}&=& g (a^{\dagger}b + b^{\dagger}a)-i\kappa a^{\dagger}a + i\kappa b^{\dagger}b\, ,\nonumber\\
  H^{(0)}_{2} &=&- i\gamma(a^{\dagger}a + b^{\dagger}b)\, ,
\end{eqnarray}
where the first $H^{\cal{PT}}_{2}$ is ${\cal{PT}}$-symmetric and it can be interpreted as a Hamiltonian of a system of equal gain and loss, both given by $\kappa.$ This time, however, the second part, i.e., $H^{(0)}_{2}$, is not just an identity operator multiplied by a constant. In this case $H^{(0)}_{2}=-i\gamma\, N$, where $N=a^{\dagger}a + b^{\dagger}b$ is the operator of the number of photons in both modes. Now it is obvious that the rate at which each of the eigenstates  scales in the EF is determined by the eigenvalues of $N$. 

The Hamiltonian~(\ref{eq:HamiltonianPT011}) in the EF is given by
\begin{eqnarray}
  \label{eq:HamiltonianPT03}
  \widetilde{H}_2&=& g (a^{\dagger}b + b^{\dagger}a)-i\kappa a^{\dagger}a + i\kappa b^{\dagger}b\, .
\end{eqnarray}
For the sake of simplicity, we assume that $g$ is real and positive.
To find the eigenvalues, we use the bosonic algebra combined with Fock space representation of~(\ref{eq:HamiltonianPT03}) \cite{Teimourpour2018}. 
To this end, we introduce the operators $[c, d]^{\rm{T}}=\boldsymbol{R}\, [a, b]^{\rm{T}}$ and 
$[c^{+}, d^{+}]^{\rm{T}}=\boldsymbol{R}\, [a^{\dagger}, b^{\dagger}]^{\rm{T}}$, where
\begin{equation}
  \label{eq:R}
\boldsymbol{R}\equiv\begin{bmatrix}
\cos\frac{\alpha}{2}&\sin\frac{\alpha}{2}\\
-\sin\frac{\alpha}{2}&\cos\frac{\alpha}{2}
\end{bmatrix}\, ,
\end{equation}
$\sin{(\alpha/2)}=\sqrt{(\lambda+i\kappa)/(2\lambda)}$, $\cos{(\alpha/2)}=\sqrt{(\lambda-i\kappa)/(2\lambda)}$ and $\lambda=\sqrt{g^2-\kappa^2}$.
Note that instead of the usual symbol "$\dagger$" we have used the symbol "+" in superscripts of the creation operators $c^{+}$ and $d^{+}$. We have changed the symbol because $c^{+}$ ($d^{+}$) is not Hermitian conjugate of $c$ ($d$)~\cite{Teimourpour2018}. Nevertheless, the operators $c$ and $d$ commute with each other and satisfy $[c,c^{+}]=1$ and $[d,d^{+}]=1$, so actually they satisfy commutation relations of independent oscillators. That is sufficient to use them for transformation of the Hamiltonian~(\ref{eq:HamiltonianPT03}) to its diagonal form
\begin{eqnarray}
  \label{eq:HamiltonianPT031}
\widetilde{H}_2&=&\lambda\,(c^{+} c-d^{+} d)\, .
\end{eqnarray}

We can immediately obtain the diagonal form of the Hamiltonian $H_2$ knowing that $H_2=H^{\cal{PT}}_{2}+H^{(0)}_{2}$, $H^{\cal{PT}}_{2}=\widetilde{H}_2$, $H^{(0)}_{2}=-i\gamma\, N$ and $N=a^{\dagger}a + b^{\dagger}b=c^{+} c+d^{+} d$:
\begin{eqnarray}
  \label{eq:HamiltonianPT032}
H_2&=&(\lambda-i\gamma)\,c^{+} c-(\lambda+i\gamma)\,d^{+} d\, .
\end{eqnarray}

Let us consider four eigenstates determined by the following excitation numbers in the modes $c$ and $d$: (1) $n_c=1$ and $n_d=0$, (2) $n_c=0$ and $n_d=1$, (3) $n_c=2$ and $n_d=0$, and (4) $n_c=0$ and $n_d=2$.
Eigenvalues of the Hamiltonian $H_2$ corresponding to these eigenstates are given by
\begin{eqnarray}
  \label{eq:HamiltonianPT033}
E_{1}=\lambda-i\gamma\, , &\,& E_{2}=-\lambda-i\gamma,\nonumber\\ 
E_{3}=2\lambda-2 i\gamma\, ,&\,& E_{4}=-2\lambda-2 i\gamma\, ,
\end{eqnarray}
whereas the eigenvalues of the Hamiltonian $\widetilde{H}_2$ are given by
\begin{eqnarray}
  \label{eq:HamiltonianPT034}
\widetilde{E}_{1}=\lambda\, , &\,& \widetilde{E}_{2}=-\lambda\, ,\nonumber\\
\widetilde{E}_{3}=2\lambda\, , &\,& \widetilde{E}_{4}=-2\lambda\, . 
\end{eqnarray}
\begin{figure}[ht]
    \centering
    \includegraphics[width=0.75\linewidth]{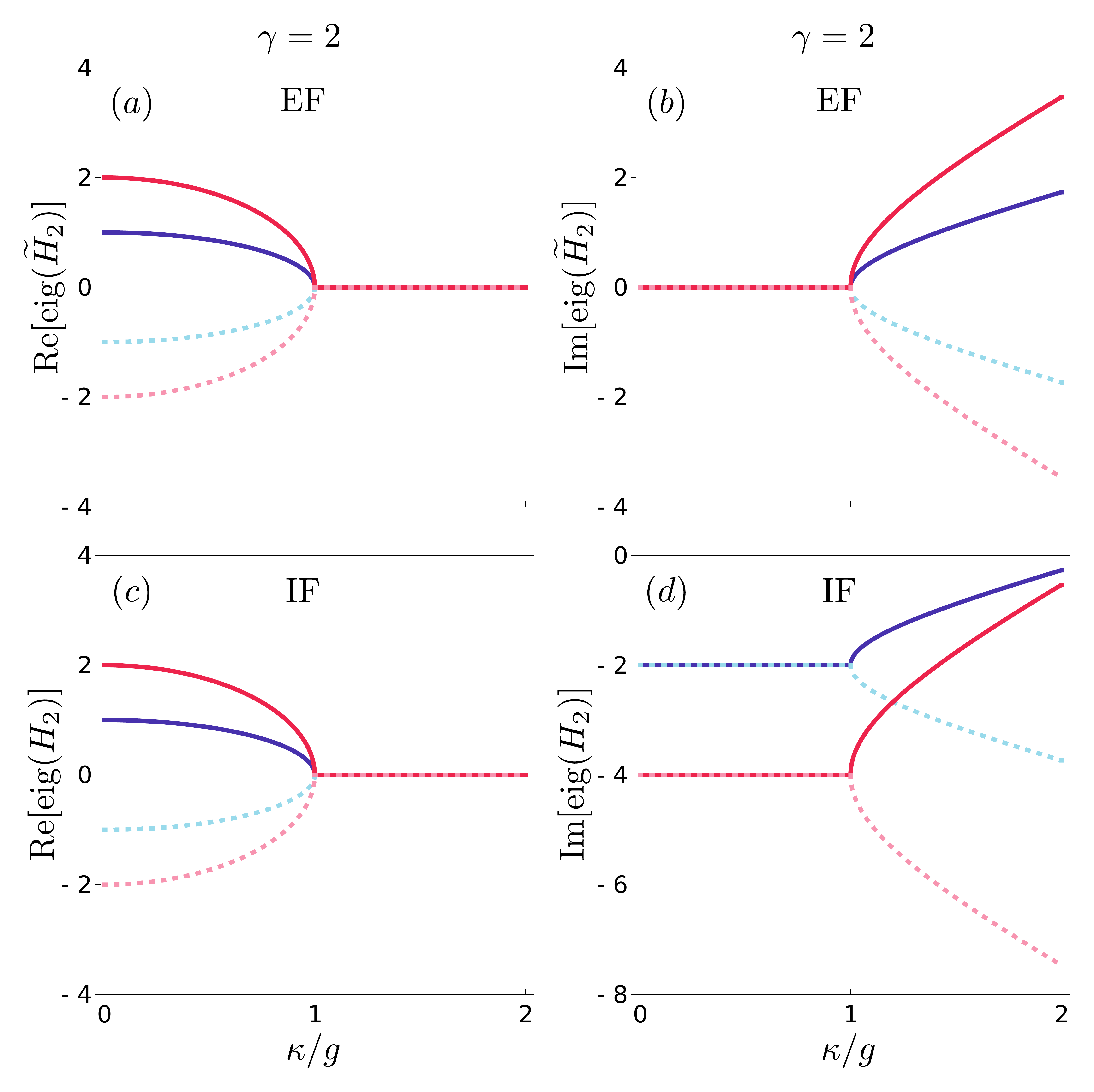}
    \caption{Eigenvalues of the Hamiltonian presented in the equilibrium frame (EF) given by Eqs.~(\ref{eq:HamiltonianPT034}) [panels (a) and (b)] and in the initial frame (IF) given by Eqs.~(\ref{eq:HamiltonianPT033}) [panels (c) and (d)] as functions of the gain/loss coefficient $\kappa$, for $g = 1$. Real parts of the eigenvalues are shown in panels (a) and (c), and imaginary parts in panels (b) and (d). Real parts are the same in both frames. Imaginary parts are different in these frames --- in the EF all eigenvalues coincide at EP, whereas in the IF coincide only those, which correspond to the same number of photons in both modes. Nevertheless, in both frames the exceptional point is the same and appears for $g=\kappa$.}
    \label{fig:passive-pt-ex1}
\end{figure}
A comparison of~(\ref{eq:HamiltonianPT033}) and~(\ref{eq:HamiltonianPT034}) is presented in Fig.~\ref{fig:passive-pt-ex1}. From this figure one can see that we can observe an EP for $\kappa=g$ in both frames, i.e., in the initial frame and in the EF. However, there are differences - in the EF all eigenvalues coincide in their real and imaginary parts (see Fig.~\ref{fig:passive-pt-ex1} panels (a) and (b)), whereas in the IF only those eigenvalues corresponding to the same number of photons $N$ coincide.
In Fig.~\ref{fig:passive-pt-ex1} one can see in panels (c) and (d) that all these eigenvalues ($E_{1}$,$E_{2}$,$E_{3}$,$E_{4}$) do not have the same real  and imaginary parts. Instead, there are two pairs of coalescing eigenvalues: \{$E_{1}$,$E_{2}$\} and \{$E_{3}$,$E_{4}$\}. One pair corresponds to eigenstates $|\psi_1\rangle=c^{+}|0\rangle_{c}|0\rangle_{d}$ and $|\psi_2\rangle=d^{+}|0\rangle_{c}|0\rangle_{d}$, for which the number of photons is $\langle\psi_1 | N |\psi_1 \rangle=\langle\psi_2 | N |\psi_2 \rangle=1$.
The second pair corresponds to $|\psi_3\rangle=c^{+\, 2}|0\rangle_{c}|0\rangle_{d}/\sqrt{2}$ and $|\psi_4\rangle=d^{+\, 2}|0\rangle_{c}|0\rangle_{d}/\sqrt{2}$, for which
$\langle\psi_3 | N |\psi_3 \rangle=\langle\psi_4 | N |\psi_4 \rangle=2$.

\subsection{Looking for hidden ${\cal{PT}}$ symmetry, in a quantum system with gain}
Let us now describe the effect of the geometric part on EP in a more interesting case, in which the Hamiltonian is given by 
\begin{eqnarray}
  \label{eq:HamiltonianPT04}
  H_3&=& g (a^{\dagger}b + b^{\dagger}a) +i\epsilon (1-i\theta) (a-a^{\dagger})
  +i\epsilon (1-i\theta^{*}) (b-b^{\dagger})
  -i\gamma_{a} a^{\dagger}a - i\gamma_{b} b^{\dagger}b-i\chi \hat{I}\, ,
\end{eqnarray}
where $\theta=\gamma(g-i\kappa)/\lambda^2$ and $\chi=2\gamma\epsilon^2/\lambda^2$.
One can see that we can express this Hamiltonian as $H_3=H^{\cal{PT}}_{3}+H^{(0)}_{3}$, where
\begin{eqnarray}
  \label{eq:HamiltonianPT05}
  H^{\cal{PT}}_{3}&=& g (a^{\dagger}b + b^{\dagger}a)+i\epsilon (a-a^{\dagger})
  +i\epsilon (b-b^{\dagger}) -i\kappa a^{\dagger}a + i\kappa b^{\dagger}b\, ,\nonumber\\
  H^{(0)}_{3} &=& \epsilon\theta (a-a^{\dagger}) + \epsilon\theta^{*} (b-b^{\dagger}) 
  - i\gamma(a^{\dagger}a + b^{\dagger}b) - i\chi \hat{I}\, .
\end{eqnarray}
In this case, the form of the geometric part linked to $H^{(0)}_{3}$ is more interesting than in the previous example. As we shall see here, the geometric part is not just the operator of the total number of photons in both modes. It is easy to check that $H^{\cal{PT}}_{3}$ is ${\cal{PT}}$-symmetric using the ${\cal{PT}}$ symmetry transformation formulas given earlier.

\begin{figure}[ht]
    \centering
    \includegraphics[width=0.75\linewidth]{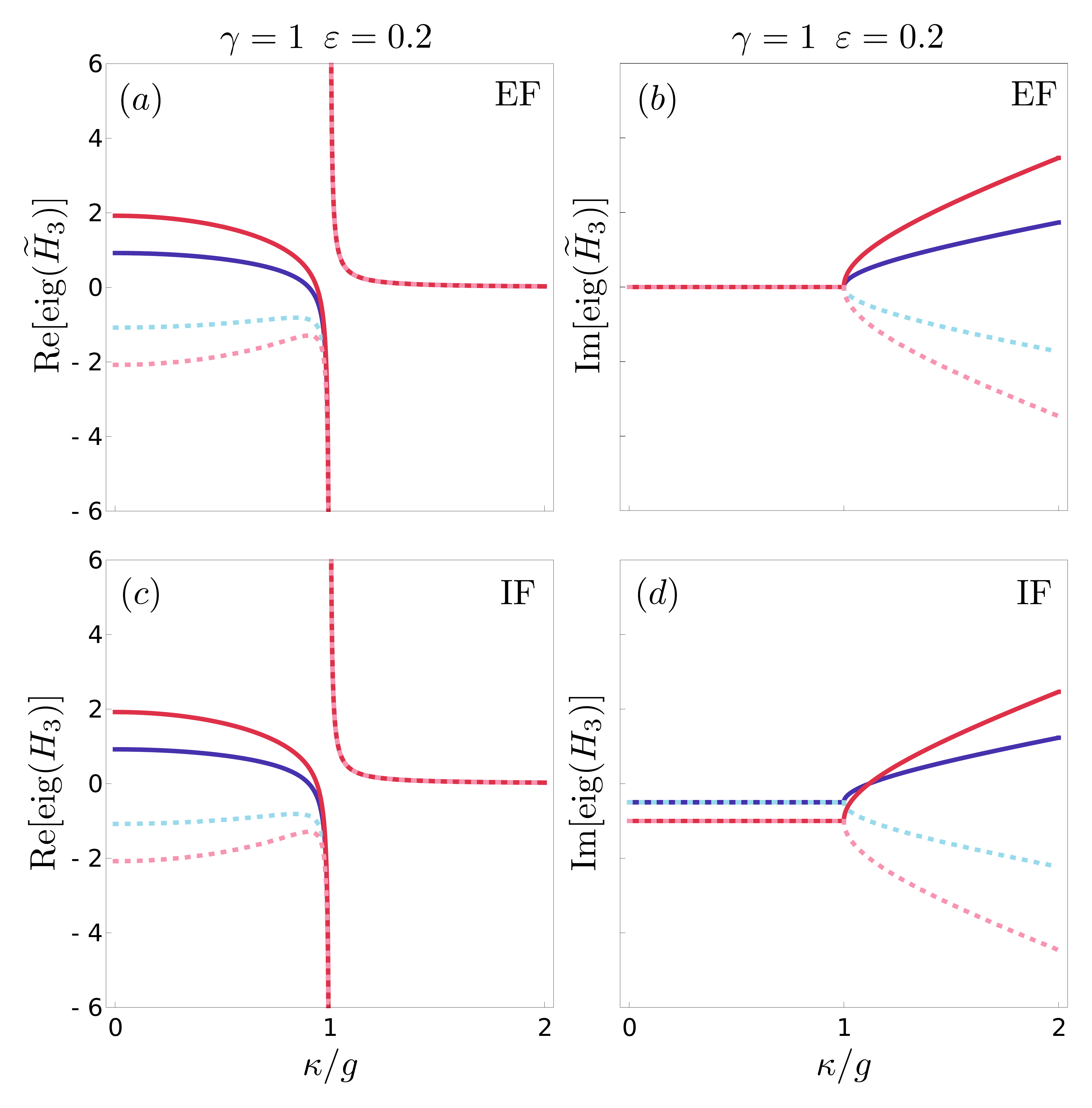}
    \caption{Eigenvalues of the Hamiltonian presented in the equilibrium frame (EF) given by Eqs.~(\ref{eq:HamiltonianPT09}) [panels (a) and (b)] and in the initial frame (IF) given by Eqs.~(\ref{eq:HamiltonianPT08}) [panels (c) and (d)] as functions of the gain/loss coefficient $\kappa$, for $g = 1$. Real parts of the eigenvalues are shown in panels (a) and (c), and imaginary parts in panels (b) and (d).}
    \label{fig:passive-pt-ex2}
\end{figure}

Let us now perform two transformations: first one defined by $[c, d]^{\rm{T}}=\boldsymbol{R}\, [a, b]^{\rm{T}}$ and $[c^{+}, d^{+}]^{\rm{T}}=\boldsymbol{R}\, [a^{\dagger}, b^{\dagger}]^{\rm{T}}$, and the second one defined by
\begin{eqnarray}
  \label{eq:CDepsOP}
c_{\varepsilon}=i c+\varepsilon_{c} \hat{I}/\lambda\, ,
&\quad& c_{\varepsilon}^{+}=-i c^{+}+\varepsilon_{c} \hat{I}/\lambda\, ,\nonumber\\
d_{\varepsilon}=i d-\varepsilon_{d} \hat{I}/\lambda\, ,
&\quad&
d_{\varepsilon}^{+}=-i d^{+}-\varepsilon_{d} \hat{I}/\lambda\, ,
\end{eqnarray}
where $\varepsilon_{c}=\varepsilon (\cos\frac{\alpha}{2}+\sin\frac{\alpha}{2})$ and $\varepsilon_{c}=\varepsilon (\cos\frac{\alpha}{2}-\sin\frac{\alpha}{2})$~\cite{Lange2020}. These operators satisfy the following commutation relations $[c_{\varepsilon},c_{\varepsilon}^{+}]=1$, $[d_{\varepsilon},d_{\varepsilon}^{+}]=1$, $[c_{\varepsilon},d_{\varepsilon}^{+}]=0$ and $[d_{\varepsilon},c_{\varepsilon}^{+}]=0$, and therefore, can be considered as annihilation and creation operators~\cite{Teimourpour2018}.
In terms of these operators the Hamiltonian takes the form $H_3=H^{\cal{PT}}_{3}+H^{(0)}_{3}$, where
\begin{eqnarray}
  \label{eq:HamiltonianPT06}
  H^{\cal{PT}}_{3}&=& \lambda\,(c_{\varepsilon}^{+} c_{\varepsilon}-d_{\varepsilon}^{+} d_{\varepsilon})
+\lambda_{0} \hat{I}\, ,\nonumber\\
  H^{(0)}_{3} &=& -i\gamma (c_{\varepsilon}^{+} c_{\varepsilon}+d_{\varepsilon}^{+} d_{\varepsilon})\, ,
\end{eqnarray}
and $\lambda_{0}=-2 g\varepsilon^2/\lambda^2$. It is seen from Eq.~(\ref{eq:HamiltonianPT06}) that the geometric part linked to $H^{(0)}_{3}$ is given by $N_{\varepsilon}=c_{\varepsilon}^{+} c_{\varepsilon}+d_{\varepsilon}^{+} d_{\varepsilon}$ and can be interpreted as the excitation number in both supermodes. After rewriting the Hamiltonian $H_3$ in terms of operators~(\ref{eq:CDepsOP}) it is easy check that $[H^{\cal{PT}}_{3},H^{(0)}_{3}]=0$, so we can expect that $H_3$ has a hidden ${\cal{PT}}$ symmetry --- $H_3$ should have EP (EPs) at the same points of parameter space as $H^{\cal{PT}}_{3}$. Of course, two eigenvalues coalesce in passive ${\cal{PT}}$-symmetric Hamiltonian only if they correspond to the same eigenvalue of $N_{\varepsilon}$. In order to check that we compare eigenvalues of $H_3$ with corresponding eigenvalues of the ${\cal{PT}}$-symmetric Hamiltonian $H^{\cal{PT}}_{3}$ for the following four eigenstates: $|\psi_1\rangle=|1\rangle_{c_{\varepsilon}}|0\rangle_{d_{\varepsilon}}$, $|\psi_2\rangle=|0\rangle_{c_{\varepsilon}}|1\rangle_{d_{\varepsilon}}$, $|\psi_3\rangle=|2\rangle_{c_{\varepsilon}}|0\rangle_{d_{\varepsilon}}$, and $|\psi_4\rangle=|0\rangle_{c_{\varepsilon}}|2\rangle_{d_{\varepsilon}}$. In the initial frame, the eigenvalues of the Hamiltonian $H_3$ corresponding to these eigenstates are given by
\begin{eqnarray}
  \label{eq:HamiltonianPT08}
E_{1}=\lambda-i\gamma+\lambda_{0}\, , &\quad&
E_{2}=-\lambda-i\gamma+\lambda_{0}\, , \nonumber \\ 
E_{3}=2\lambda-2 i\gamma+\lambda_{0}\, , &\quad&
E_{4}=-2\lambda-2 i\gamma+\lambda_{0}\, .
\end{eqnarray}
In EF, the Hamiltonian is given by $\widetilde{H}_3$ and it has the following eigenvalues
\begin{eqnarray}
  \label{eq:HamiltonianPT09}
\widetilde{E}_{1}=\lambda+\lambda_{0}\, , &\quad&
\widetilde{E}_{2}=-\lambda+\lambda_{0}\, , \nonumber \\
\widetilde{E}_{3}=2\lambda+\lambda_{0}\, , &\quad&
\widetilde{E}_{4}=-2\lambda+\lambda_{0}\, .
\end{eqnarray}
In Fig.~\ref{fig:passive-pt-ex2} these eigenvalues are plotted as functions of $\kappa$ for $g=1$.
One can see that Hamiltonians describing the quantum system in IF and EF have a spectral singularity at $\kappa = g$. If $\kappa\to g$ then $\lambda\to 0$ and all these eigenvalues in the ${\cal{PT}}$-symmetric case (i.e., in the EF, where the Hamiltonian is given by $\widetilde{H}_3$) tend to one value. In the passive case ${\rm{Im}}(E_{1})$ and ${\rm{Im}}(E_{2})$ tend to $-i\gamma$, while ${\rm{Im}}(E_{3})$ and ${\rm{Im}}(E_{4})$ tend to $-2 i\gamma$, because $\langle \psi_1|N_{\varepsilon}|\psi_1\rangle=\langle \psi_2|N_{\varepsilon}|\psi_2\rangle=1$ and $\langle \psi_3|N_{\varepsilon}|\psi_3\rangle=\langle \psi_4|N_{\varepsilon}|\psi_4\rangle=2$.

\section{Discussion}

We have considered passive systems described by non-Hermitian Hamiltonians, which are not ${\cal{PT}}$-symmetric. We have shown that if non-${\cal{PT}}$-symmetric Hamiltonians can be expressed as a sum of two terms, a ${\cal{PT}}$-symmetric term and a second term commuting with the first one; such Hamiltonian have hidden ${\cal{PT}}$ symmetry. Thus, it can display exceptional points. This is a consequence of the fact that such a non-${\cal{PT}}$-symmetric Hamiltonian can be transformed from its initial frame (IF) to a apparently ${\cal{PT}}$-symmetric form expressed in an equilibrium frame (EF). Hence, rewriting the Hamiltonian in EF reveals its hidden ${\cal{PT}}$ symmetry.

We have also shown that the second part of the Hamiltonian given in IF plays an important role in the EF --- it determines the scaling rate (i.e., the exponential decay  rate of a given eigenstate). Therefore, there is a difference between EF and the IF, i.e., the number of eigenvalues, which coincide at exceptional points is usually greater in the EF than in the IF. In IF, the eigenvalues coincide only when they correspond to the same value of the second part.

It is worth to note that the presented here knowledge might be helpful to investigate ${\cal{PT}}$-symmetric systems experimentally. Researchers discover interesting phenomena frequently assuming non-Hermitian Hamiltonians, which have the gain and the damping terms. However, contrary to the damping, it is not easy to realise the incoherent gain. In passive ${\cal{PT}}$-symmetric systems the incoherent gain is not involved. One can add to a ${\cal{PT}}$-symmetric Hamiltonian a proper term commuting with it to obtain a passive ${\cal{PT}}$-symmetric Hamiltonian.

\section*{Acknowledgements}
This work was supported by the Polish National Science Centre (NCN) under the
Maestro Grant No. DEC-2019/34/A/ST2/00081.

%\bibliographystyle{apsrev}
%\bibliography{frame}

\end{document}